\documentclass[aps,prd,amsmath,two column,nofootinbib,amssymb,showpacs]{revtex4}
\usepackage{amssymb}
\usepackage{txfonts}
\usepackage{amsfonts}
\usepackage{mathrsfs}
\usepackage{epsfig,bm,dcolumn}
\usepackage{graphicx}
\usepackage{color}
\usepackage{amsmath}
\usepackage{dcolumn}
\usepackage{overpic}
\usepackage{slashed}

\begin{document}
\title{Solitonic modulation and Lifshitz point in an external magnetic field within Nambu--Jona-Lasinio model}

\author{Gaoqing Cao and Anping Huang}
\affiliation{Department of Physics, Tsinghua University and Collaborative Innovation Center of Quantum Matter, Beijing 100084, China}

\date{\today}
\begin{abstract}
 We study the inhomogeneous solitonic modulation of a chiral condensate within the effective Nambu--Jona-Lasinio model when a constant external magnetic field is present. The self-consistent Pauli-Villars regularization scheme is adopted to manipulate the ultraviolet divergence encountered in the thermodynamic quantities. In order to determine the chiral restoration lines efficiently, a new kind of Ginzburg-Landau expansion approach is proposed here. At zero temperature, we find that both the upper and lower boundaries of the solitonic modulation oscillate with the magnetic field in the $\mu$--$B$ phase diagram which is actually the de Hass-van Alphan (dHvA) oscillation. It is very interesting to find out how the tricritical Lifshitz point $(T_L,\mu_L)$ evolves with the magnetic field: There are also dHvA oscillations in the $T_L$--$B$ and $\mu_L$--$B$ curves, though the tricritical temperature $T_L$ increases monotonically with the magnetic field.
\end{abstract}


\maketitle
\section{Introduction}
The inhomogeneous Larkin-Ovchinnikov-Fulde-Ferrell (LOFF) state has attracted a lot of interests since its proposition in 1960s~\cite{FFLO}. Though no direct evidence of LOFF state has been discovered in experiments yet, recent developments of ultracold atomic physics might provide a good opportunity to pin down the exotic state because of its strong controllability~\cite{Zheng2013,Shenoy2013,Wu2013}. In condensed matter physics, an external magnetic field will split the single-particle dispersion of electrons with spin-up and spin-down through Zeeman effect and a large mismatch of the Fermi surfaces usually favors the pairing with finite momentum~\cite{FFLO}. While in quantum chromodynamic (QCD) systems, baryon chemical potential directly plays a mismatch between quark and anti-quark and isospin chemical potential plays a mismatch between u and d quarks; the sources of LOFF state are very rich in quark matter and nuclear matter~\cite{Nakano:2004cd,Nickel:2009ke,Nickel:2009wj,Carignano:2010ac,Carignano:2014jla,He:2006tn,Abuki:2013pla,Abuki:2014bya,Bowers:2002xr,
Rajagopal:2006ig,He:2006vr,Nickel:2008ng,Heinz:2013hza,Buballa:2014tba,Cao:2015rea}.

Specially, in the study of chiral symmetry breaking and restoration, it was found that the original first-order transition line in the $T$--$\mu$ phase diagram~\cite{Zhuang:1994dw} would be covered by the inhomogeneous phase with the solitonic modulation (SM) or dual chiral density wave (DCDW) modulation chiral condensate~\cite{Nickel:2009wj,Carignano:2014jla}. But different models give different predictions about which inhomogeneous phase is much more favored. A recent study in the renormalizable quark-meson model suggested that the solitonic modulation usually has lower free energy and the phase boundaries are both of second order which indicates the existence of a tricritical Lifshitz point~\cite{Carignano:2014jla}. On the other hand, at the presence of a constant magnetic field, the features of chiral symmetry breaking and restoration with LOFF state alter a lot~\cite{Frolov:2010wn,Tatsumi:2013nga}. Because of the asymmetry between the particle and antiparticle dispersions
at the lowest Landau level (LLL), the DCDW modulation was found to be much more favored at non-vanishing chemical potential~\cite{Frolov:2010wn} and the transition point was found to be a tricritical point at vanishing
chemical potential~\cite{Tatsumi:2013nga}. However, due to the sensitivity of the inhomogeneous state to the choice of regularizations in the chiral effective Nambu--Jona-Lasinio (NJL) model~\cite{Frolov:2010wn,Meissner:1990se} and
the ambiguous "intermediate regularization" introduced in Ref.~\cite{Frolov:2010wn}, the real situation is still unclear.

In this work, we explore the features of chiral symmetry breaking and restoration in a constant external magnetic field by taking
the solitonic modulation chiral condensate into account. The main advantage is that this situation can be handled self-consistently
 by adopting the Pauli-Villars (PV) regularization scheme. Besides, in recent years, inverse magnetic catalysis (IMC) effect was found in QCD system
~\cite{Bali2012} and inspired a lot of discussions relevant to the effects of magnetic field
~\cite{Bruckmann:2013oba,Fukushima:2012kc,Chao:2013qpa,Yu:2014sla,Cao:2014uva,Ferreira:2013tba,Ferrer:2014qka,Farias:2014eca,Mueller:2015fka,Preis:2010cq,Cao:2015xja,Miransky:2015ava}.
Experts mainly attributed this anomalous phenomenon to the asymptotic freedom or earlier deconfinement transition~\cite{Bruckmann:2013oba,Ferreira:2013tba,Farias:2014eca,Miransky:2015ava}. Nevertheless, at very large magnetic field, it is possible that the deconfinement transition might happen earlier than chiral symmetry restoration with increasing temperature~\cite{Fraga:2010qb,Miransky:2015ava}. In the case of finite chemical potential, it is usually thought that deconfinement and chiral symmetry restoration transition separates
with each other which then gives rise to the quarkyonic matter~\cite{McLerran}; thus, if the IMC effect holds
in this case is still unknown, especially near the Lifshitz point. For simplicity, we still adopt the initial Nambu--Jona-Lasinio model and introduce the magnetic field through covariant derivative.
We'll give some comments whenever the IMC effect should be considered.

The paper is organized as following: In Sec.\ref{Formulism}, we give a general theoretical framework for the study of chiral symmetry
breaking and restoration with Sec.\ref{soliton} presenting the formulism for solitonic modulation in NJL model and Sec.\ref{GL} discussing a new kind of Ginzburg-Landau (GL)
expansion approach. The numerical results will be shown in Sec.\ref{results} and we finally summarize in Sec.\ref{summary}.
\section{Theoretical framework}\label{Formulism}
\subsection{Nambu--Jona-Lasinio model and solitonic modulation}\label{soliton}
The Lagrangian of Nambu--Jona-Lasinio model in a constant external magnetic field is
\begin{eqnarray}\label{NJL}
{\cal L}=\bar{\psi}(i\slashed{D}-m_0-\mu\gamma^0)\psi
+{G}\left[(\bar{\psi}\psi)^{2}+(\bar{\psi}i\gamma_{5}{\boldsymbol\tau}\psi)^{2}\right],
\end{eqnarray}
where $\psi(x)=(u(x),d(x))^T$ denotes the two-flavor quark field with color degrees of freedom $N_c=3$ and $D_\mu=\partial_\mu+iQA_\mu$ is the covariant derivative with the subscripts $\mu=1,2,3$ corresponding to the coordinates ${\rm x,y,z}$ and the charge matrix $Q=\text{diag}(q_{\rm u},q_{\rm d})=\text{diag}({2e/3},-{e/3})$ in flavor space. Without loss of generality, we set the magnetic field to be along z-direction and the vector potential to be expressed with Landau gauge, that is, $A_{\mu}=(0,0,-B{\rm x},0)\;(B>0)$ . Here, $m_0$ is the current mass of u and d quarks which we set to zero for simplicity as it will not affect the qualitative results~\cite{Nickel:2009wj}, $\mu$ is quark chemical potential and ${\boldsymbol\tau}=(\tau^1,\tau^2,\tau^3)$ are Pauli matrices in flavor space. For the convenience of later discussions, we'd like to choose the explicit forms of $\gamma$-matrices in Weyl representation \cite{Peskin1995}:
\begin{eqnarray}\label{gamma}
\gamma^0=\left(\begin{array}{cc} 0& I\\ I& 0\end{array}\right),\ \ \
\gamma^i=\left(\begin{array}{cc} 0& \sigma^i\\ -\sigma^i&0\end{array}\right),
\end{eqnarray}
where $\sigma^1,\sigma^2,$ and $\sigma^3$ are Pauli matrices and $I$ is the $2\times2$ identity matrix in Dirac spinor space. As well known, in chiral limit $m_0=0$, the Lagrangian has ${\rm SU_ L}(2)\times {\rm SU_ R}(2)$ chiral symmetry with the left-handed and right-handed transformations $\psi(x)\rightarrow\exp(i{(1\mp\gamma_5)}\boldsymbol{\tau\cdot\theta}/2)\;\psi(x)$, respectively. If the expectation value of $\bar{\psi}\psi$ or $\bar{\psi}i\gamma_{5}{\boldsymbol\tau}\psi$ is nonvanishing, it is easy to see that the symmetry of the Lagrangian will be spontaneously broken. Thus, $\bar{\psi}\psi$ and $\bar{\psi}i\gamma_{5}{\boldsymbol\tau}\psi$ are both
order parameters for chiral symmetry breaking and restoration.

In order to study the expectation values of the order parameters, we introduce four auxiliary fields $\sigma(x)$ and ${\boldsymbol\pi}(x)$. Then the original partition function ${\cal Z}=\int[{\cal D}\bar{\psi}][{\cal D}\psi]e^{i{\cal L}}$ can be changed to the following form:
\begin{eqnarray}
{\cal Z}&=&\int[{\cal D}\bar{\psi}][{\cal D}\psi][{\cal D}\sigma][{\cal D}\pi]\exp\Big\{i\int d^4x \Big[-{1\over 4G}(\sigma^2+\mathbf{\pi}^2)\nonumber\\
&&+\bar{\psi}(i\slashed{D}-\sigma-i\gamma_5\boldsymbol{\pi\cdot\tau}-\mu\gamma^0)\psi
\Big]\Big\},
\end{eqnarray}
which is actually the Hubbard-Stratonovich transformation~\cite{HS}. Then, by completing the functional integration over the quark degrees of freedom, the model can be bosonized with the partition function given as
\begin{eqnarray}\label{partition}
{\cal Z}&=&\int[{\cal D}\sigma][{\cal D}\pi]\exp\Big\{-i\Big[\frac{1}{4G}\int d^4x(\sigma^2+\mathbf{\pi}^2)\nonumber\\
&&+iN_{\rm c}{\rm Tr\ln}(i\slashed{D}-\sigma-i\gamma_5\mathbf{\pi\cdot\tau}-\mu\gamma^0)\Big]\Big\},
\end{eqnarray}
where the trace is taken over the coordinate, spinor and flavor spaces. In mean field approximation, let's set $\langle\sigma(x)\rangle=m_{\rm r}(x)$, $\langle\pi^3(x)\rangle=m_{\rm i}(x)$, and $\langle\pi^1(x)\rangle=\langle\pi^2(x)\rangle=0$. This setting is general when the magnetic field is absent because of the rotational symmetry in the flavor space. However, when a finite magnetic field is present, other cases with $\langle\pi^1(x)\rangle$ or $\langle\pi^2(x)\rangle$ nonvanishing are quite different from this. In principal, these cases are hard to evaluate precisely because of the electric charges carried by $\pi^1$ and $\pi^2$. According to the lattice QCD (LQCD) calculation~\cite{Endrodi:2014lja} and the Ginzburg-Landau analysis of our previous work~\cite{Cao:2015xja}, $\pi^1(\pi^2)$ condensation is just like superconductor and the magnetic field disfavors the pion superfluidity, so these cases can be safely neglected here. Then, the thermodynamic potential for the chosen setting can be evaluated as
\begin{eqnarray}
\Omega&=&{T\over V}\big[-N_{\rm c}{\rm Tr\ln}\big(i\slashed{D}-m_{\rm r}(x)-i\gamma_5{\tau^3}m_{\rm i}(x)-\mu\gamma^0\big)\nonumber\\
&&+\frac{1}{4G}\int d^4x\big(m_{\rm r}^2(x)+m_{\rm i}^2(x)\big)\big]\nonumber\\
&=&{T\over V}\big[\!-N_{\rm c}{\rm Tr\ln}\big(i\slashed{D}\!-\!{1\!+\!\gamma^5{\tau^3}\over2}m(x)\!-\!{1\!-\!\gamma^5{\tau^3}\over2}m^*(x)\!-\!\mu\gamma^0\big)\nonumber\\
&&+\frac{1}{4G}\int d^4x|m(x)|^2\big],
\end{eqnarray}
where the mass gap $m(x)=m_{\rm r}(x)+i\,m_{\rm i}(x)$, and we work in Euclidean space with the integral of the imaginary time $x_4=ix_0$ in the region $[0,1/T]$.

In order to obtain an explicit expression for $\Omega$, it's essential to evaluate the contribution of quarks. Let's suppose that the inhomogeneous mass gap is one-dimensional and along the magnetic field, that is, $m(x)=m({\rm z})$, as the Taylor expansion analysis indicated that the inhomogeneity along other dimensions is disfavored~\cite{Frolov:2010wn}. Then, the most important mission left is to evaluate the eigenvalues of the following Hamiltonian:
\begin{eqnarray}\label{Hf}
H_{\rm f}&=&-\gamma^0\big[i\slashed{\boldsymbol D}_{\rm f}-{1+\gamma^5\over2}m({\rm z})-{1-\gamma^5\over2}m^*({\rm z})\big]\nonumber\\
&=&-\gamma^0\big[i\gamma^1\partial_{\rm x}+i\gamma^2(\partial_{\rm y}-iq_{\rm f}B{\rm x})+i\gamma^3\partial_{\rm z}\nonumber\\
&&-{1+\gamma^5\over2}m({\rm z})-{1-\gamma^5\over2}m^*({\rm z})\big],
\end{eqnarray}
where the subscript ${\rm f}$ stands for the flavor $u$ or $d$. If we expand the spinors with Ritus's method by separating variables, the first two terms in the square bracket give rise to a Landau level term $i\sqrt{2n |q_{\rm f}B|}\gamma^2$~\cite{Ritus,Warringa:2012bq}. The left terms actually correspond to a one-dimensional NJL model or Gross-Neveu (GN) model with the following Hamiltonian:
\begin{eqnarray}
H_{\rm GN}&=&-\gamma^0\big[i\gamma^3\partial_{\rm z}-{1+\gamma^5\over2}m({\rm z})-{1-\gamma^5\over2}m^*({\rm z})\big]\nonumber\\
&=&\qquad\left(\begin{array}{cccc}  i\partial_{\rm z}&0&m({\rm z})&0\\ 0&-i\partial_{\rm z}&0&m({\rm z})\\m^*({\rm z})&0&-i\partial_{\rm z}&0\\0&m^*({\rm z})&0&i\partial_{\rm z}\end{array}\right).
\end{eqnarray}
The Hamiltonian can be brought to a block diagonal form by taking a similitude transformation, that is,
\begin{eqnarray}
U^{-1}H_{\rm GN}U=\left(\begin{array}{cc}  H_{{\rm z}}\big(m({\rm z})\big)&0\\ 0&H_{\rm z}\big(m^*({\rm z})\big)\end{array}\right),
\end{eqnarray}
where the involved matrices are respectively:
\begin{eqnarray}
H_{\rm z}\big(m({\rm z})\big)=\left(\begin{array}{cc}  i\partial_{\rm z}&m^*({\rm z})\\ m({\rm z})&-i\partial_{\rm z}\end{array}\right),U=\left(\begin{array}{cccc}  1&0&0&0\\ 0&0&0&1\\0&1&0&0\\0&0&1&0\end{array}\right).
\end{eqnarray}

For a given inhomogeneous state, the explicit form of the thermodynamic potential usually can be evaluated with the help of the density of states. However, one should be cautious when $m({\rm z})$ is not real. In this case, the spinors are actually half valid at the LLL and take the forms $u(x)=\big(u_1(x),0,u_3(x),0\big)^T$ and $d(x)=\big(0,d_2(x),0,d_4(x)\big)^T$, respectively~\cite{Warringa:2012bq}. Then the spectra $\{\varepsilon\}$ of $H_{\rm GN}$ are not symmetric with $\{-\varepsilon\}$. Take the DCDW modulation ($m({\rm z})=me^{2ik{\rm z}}$) of $u$ quark for example, the spectra are $\{\pm\sqrt{p_{\rm z}^2+m^2}+k\}$. As has been mentioned, these sign asymmetric spectra make the regularization very difficult at finit chemical potential due to the non-renormalizable nature of NJL model~\cite{Frolov:2010wn,Meissner:1990se}. For the solitonic modulation, $m({\rm z})$ is real and it can be checked that the LLL spectra are sign symmetric. Thus, this case can be treated self-consistently.

For the solitonic modulation, the mass gap takes the following form~\cite{Schnetz:2005ih}:
 \begin{eqnarray}
M(m,\nu,{\rm z})&=&m\Big(\nu\,{\rm sn}(\mathbf{K}(\nu)|\nu)\,{\rm sn}(m {\rm z}|\nu)\,{\rm sn}(m {\rm z}+\mathbf{K}(\nu)|\nu)\nonumber\\
&&+\frac{{\rm cn}(\mathbf{K}(\nu)|\nu)\,{\rm dn}(\mathbf{K}(\nu)|\nu)}{{\rm sn}(\mathbf{K}(\nu)|\nu)}\Big),
\end{eqnarray}
where ${\rm sn}, {\rm cn}$ and ${\rm dn}$ are elliptic Jocobi functions with elliptic modulus $\sqrt{\nu}$. And the thermodynamic potential can be derived straightforwardly from the case with vanishing magnetic field~\cite{Nickel:2009wj} because the transversal degrees of freedom are irrelevant to the longitudinal one. By replacing the transverse momenta with the Landau Levels, the thermodynamic potential can be expressed explicitly as
\begin{eqnarray}
\Omega(T,\mu,B;m,\nu)&=&\!\!{1\over4GL}\int_0^L\!\!M^2(m,\nu,z){\rm d}z\!-\!\!\!\sum_{q=q_{\rm u},q_{\rm d}}\!\!N_c{|qB|\over2\pi}\sum_{n=0}\alpha_n\nonumber\\
&&\int_0^\infty d\varepsilon\;\rho(\varepsilon;m,\nu)f(T,\mu,n,q,B,\varepsilon),
\end{eqnarray}
where $L=2\mathbf{K}(\nu)/m$ is the period of $M(m,\nu,z)$, $\alpha_n=2-\delta_{\rm n,0}$ stands for the degeneracy of the $n$-th Landau level, and the integrand is
\begin{eqnarray}
f(T,\mu,n,q,B,\varepsilon)&=&\epsilon(n,q,B,\varepsilon)+T\ln(1+e^{-(\epsilon(n,q,B,\varepsilon)-\mu)/T})\nonumber\\
&&+T\ln(1+e^{-(\epsilon(n,q,B,\varepsilon)+\mu)/T}),
\end{eqnarray}
with the excitation energy $\epsilon(n,q,B,\varepsilon)=\big(2n|qB|+\varepsilon^2\big)^{1/2}$. The corresponding density of states for solitonic modulation is given by~\cite{Schnetz:2004vr}
\begin{eqnarray}
\rho(\varepsilon;m,\nu)&=&{1\over\pi}\frac{\varepsilon^2-m^2\mathbf{E}(\nu)/\mathbf{K}(\nu)}
{\sqrt{(\varepsilon^2-m^2)\big(\varepsilon^2-(1-\nu)m^2\big)}}\nonumber\\
&&\big[\theta(\varepsilon^2-m^2)-\theta\big(-\varepsilon^2+(1-\nu)m^2\big)\big],
\end{eqnarray}
where $\mathbf{K}(\nu)$ is the quarter period, $\mathbf{E}(\nu)$ is the incomplete elliptic integral and $\theta(x)$ is the step function. For the convenience of numerical calculations, the integral in the first part of $\Omega$ can be worked out presicely to give \cite{Schnetz:2005ih}
\begin{eqnarray}
M^2(m,\nu)=m^2\Big({1\over{\rm sn}^2(\mathbf{K}(\nu)|\nu)}-{2\mathbf{E}(\nu)\over\mathbf{K}(\nu)}+1-\nu\Big).
\end{eqnarray}
The second part is divergent and we refer to PV regularization scheme as it is much softer than others and can avoid artifacts when magnetic field is present~\cite{Cao:2015xja}. Then the convergent form of the thermodynamic potential is
\begin{eqnarray}
\Omega(T,\mu,B;m,\nu)&=&{M^2(m,\nu)\over4G}-\sum_{q=q_{\rm u},q_{\rm d}}N_c{|qB|\over2\pi}\sum_{n=0}\alpha_n\nonumber\\
&&\!\!\!\int_0^\infty \!\!\!d\varepsilon\;\rho(\varepsilon;m,\nu)f_{\rm PV}(T,\mu,n,q,B,\varepsilon),
\end{eqnarray}
where $f_{\rm PV}(T,\mu,n,q,B,\varepsilon)=\sum_{i=0}^3c_{\rm i}f(T,\mu,n,q,B,\sqrt{\varepsilon^2+i\Lambda^2})$ with $c_0=-c_3=1, c_1=-c_2=-3$. The advantages of the PV regularization scheme are ready to see: In the limit $\nu\rightarrow0$ or $m\rightarrow0$, we can reproduce the PV regularized thermodynamic potential for the chiral symmetry restoration phase $(\chi SR)$ as it should be. In the limit $\nu\rightarrow1$, the PV regularized thermodynamic potential for the homogeneous chiral symmetry breaking phase $(\chi SB)$ can also be reproduced. Thus, the PV regularization scheme guarantees the self-consistency of solitonic modulation in several limits. Finally, the ground state should be determined by minimizing $\Omega(T,\mu,B;m,\nu)$ with respect to $m$ and $\nu$ for given parameters $T,\mu$ and $B$. And phase transition happens with the change of parameters: When $\nu$ changes from $1$ to a smaller one, the system transits from $\chi SB$ phase to SM phase; when it changes from $0<\nu<1$ to 0, the system transits from SM phase to $\chi SR$ phase.
\subsection{Ginzburg-Landau expansion with small $\nu$}\label{GL}
In order to evaluate the chiral symmetry restoration transition and the Lifshitz point more efficiently, we'd like to introduce a different Ginzburg-Landau expansion scheme compared to that of Ref.~\cite{Nickel:2009wj}, that is, expand the thermodynamic potential $\Omega(T,\mu,B;m,\nu)$ with respect to the elliptic modulus $\nu$. The Taylor expansions of $M^2(m,\nu)$ and $\rho(\varepsilon;m,\nu)$ around small $\nu$ are respectively
\begin{eqnarray}
M^2(m,\nu)&=&m^2\big(\sqrt{\nu}-{7\nu\over8}+o(\nu^{3/2})\big),\\
\rho(\varepsilon;m,\nu)&=&{1\over\pi}+{m^2\sqrt{\nu}\over2\pi(\varepsilon^2-m^2)}-{7m^2\nu\over16\pi(\varepsilon^2-m^2)}\nonumber\\
&&-{m\nu\over2\pi}\delta(\varepsilon-m)+o(\nu^{3/2}),\label{rho}
\end{eqnarray}
where $o(\nu^{3/2})$ is the Peano form of the remainder and the third term in Eq.(\ref{rho}) is from the step functions. Thus, the thermodynamic potential has the following form
\begin{eqnarray}
\Omega(T,\mu,B;m,\nu)&=&\Omega(T,\mu,B;m,0)+m^2\beta(T,\mu,B;m)\sqrt{\nu}\nonumber\\
&&+\Big(-{7\over8}m^2\beta(T,\mu,B;m)+\gamma(T,\mu,B;m)\Big)\;\nu\nonumber\\
&&+o(\nu^{3/2}),
\end{eqnarray}
\begin{eqnarray}
&&\!\!\!\!\!\!\!\!\!\!\!\!\!\!\!\beta(T,\mu,B;m)={1\over4G}-\sum_{q=q_{\rm u},q_{\rm d}}N_c{|qB|\over2\pi}\sum_{n=0}\alpha_n\int_0^\infty d\varepsilon\nonumber\\
&&\ \ \ \ \ \ \ \ \quad\;{1\over2\pi(\varepsilon^2-m^2)}f_{\rm PV}(T,\mu,n,q,B,\varepsilon),\\
&&\!\!\!\!\!\!\!\!\!\!\!\!\!\!\!\gamma(T,\mu,B;m)=m\!\!\!\sum_{q=q_{\rm u},q_{\rm d}}\!N_c{|qB|\over(2\pi)^2}\sum_{n=0}\alpha_nf_{\rm PV}(T,\mu,n,q,B,m).
\end{eqnarray}
\indent Note that $\Omega(T,\mu,B;m,0)$ is just the thermodynamic potential for $\chi SR$ phase which doesn't depend on $m$. However, $m^2\beta(T,\mu,B;m)$ is mass dependent and the value of $m$ should be determined by minimizing this coefficient for given parameters which gives the lowest free energy around $\nu\sim0$. The integral over $\varepsilon$ should be understood as Cauchy Principal value integration and then $\beta(T,\mu,B;m)$ is real and convergent. The minimum of $\beta(T,\mu,B;m)$ directly determines which phase is the system in: If the minimum is negative, the $\chi SB$ phase or SM phase is more favored; if positive, the $\chi SR$ phase is more favored; and $\min\big(\beta(T,\mu,B;m)\big)=0$ just determines the chiral symmetry restoration point. At the transition point, the next-order coefficient reduces to $\gamma(T,\mu,B;m)$ which is semi-positive definite (only equals to zero when $m=0$) and this means that the phase transition is always of second order. It should be clarified that a nonzero expectation value of $m$ around $\nu\sim0$ doesn't necessarily mean the transition cannot be second order, because what really matters is $M(m,\nu,{\rm z})$ which of course vanishes at $\nu=0$.

Furthermore, this kind of GL expansion approach is also capable to find the Lifshitz point where the solution with $\nu=0$ is consistent with the solution $\nu=1$. This can happen only when the expectation value of $m$ is zero and the Lifshitz point is in fact the critical point between $m=0$ and $m\neq0$. Nevertheless, there is a small defect with this approach: The derivative of the coefficient $\beta(T,\mu,B;m)$ with respect to $m$ is divergent around the integral region $\varepsilon\sim m$ as can be seen from the following
\begin{eqnarray}
{\partial\beta(T,\mu,B;m)\over m\;\partial m}&=&\!\!-\!\!\!\!\sum_{q=q_{\rm u},q_{\rm d}}\!\!\!\!N_c{|qB|\over2\pi}\!\!\sum_{n=0}\alpha_n\!\int_0^\infty\!\!\! d\varepsilon{f_{\rm PV}(T,\mu,n,q,B,\varepsilon)\over\pi(\varepsilon^2-m^2)^2}.\nonumber\\
\end{eqnarray}
Therefore, the minimum can only be evaluated by direct scanning of $\beta(T,\mu,B;m)$ over $m$ instead of a new kind of "gap equation".
\section{The phase diagrams and Lifshitz point}\label{results}
As had already been illuminated in the quark-meson (QM) model \cite{Nickel:2009wj}, the expectation value of $m$ in vacuum affects the qualitative results quite much about the existence of the solitonic modulation. In order to show the results explicitly, we choose $m=330~\text{MeV}$ as a moderate choice and keep the pion decay constant $f_{\rm \pi}$ to the experimental value $93~\text{MeV}$. Then, according to the following relations~\cite{Klevansky:1992qe}:
\begin{eqnarray}
\langle\bar{\psi}\psi\rangle&\equiv&-{m\over2G}=-{6m\over4\pi^2}\sum_{i=0}^3c_i(m^2+i\Lambda^2)\ln{m^2+i\Lambda^2\over m^2},\\
f_{\rm \pi}^2&=&{N_{\rm c}m^2\over4\pi^2}\sum_{i=0}^3c_i\ln{m^2+i\Lambda^2\over m^2},
\end{eqnarray}
the parameters of the PV regularized NJL model can be fixed as $\Lambda=0.786~\text{GeV}$ and $G\Lambda^2=6.24$. This corresponds to a chiral condensate $\langle\bar{u}u\rangle=\langle\bar{d}d\rangle=\langle\bar{\psi}\psi\rangle/2=-(0.20~\text{GeV})^3$ in the vacuum which is a little smaller than the LQCD result $\langle\bar{u}u\rangle=\langle\bar{d}d\rangle=-(0.25~\text{GeV})^3$.

Both the NJL model and QM model predicted that the transitions from $\chi SB$ phase to SM phase and from SM phase to $\chi SR$ phase are both of second order in the absence of magnetic field~\cite{Nickel:2009wj,Carignano:2014jla}. We carefully check the case with a constant magnetic field and find it remains the same: As there is only one minimum of $\Omega(T,\mu,B;m,\nu)$ with respect to $\nu$ which alone determines the transition points for given parameters and nonzero $m$ and the conventional first-order transition from $\chi SB$ phase to $\chi SR$ phase lies between these two, the transitions must both be of second order.

\begin{figure}[!htb]
\centering
\includegraphics[width=0.45\textwidth]{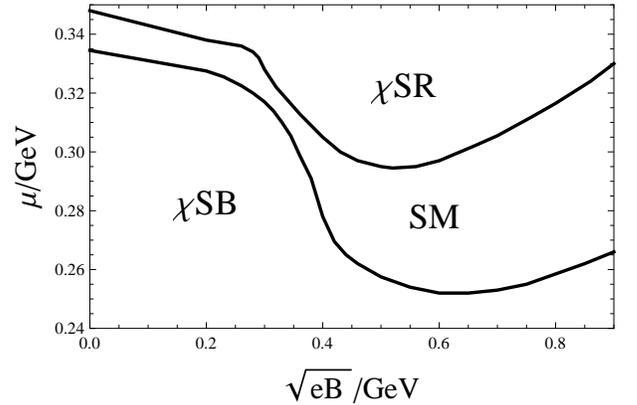}
\caption{The $\mu$--$B$ phase diagram at zero temperature. The upper and lower boundaries correspond to the transition from solitonic modulation phase to chiral symmetry restoration phase and from chiral symmetry breaking phase to solitonic modulation phase, respectively.}\label{magnetic}
\end{figure}

In order to explore the effect of magnetic field, we'd like to start firstly at zero temperature. In this case, the integrand becomes
\begin{eqnarray}
f(0,\mu,n,q,B,\varepsilon)={1\over2}\sum_{s=\pm}|\epsilon(n,q,B,\varepsilon)-s\mu|,
\end{eqnarray}
and the numerical results for the upper and lower boundaries of SM phase are shown in Fig.\ref{magnetic}. It is interesting to find that the critical chemical potential $\mu$ oscillates with the magnetic field $B$ at both boundaries which is actually an illumination of the dHvA effect~\cite{dHvA}. This is consistent with that found in the study of homogeneous chiral condensate~\cite{Cao:2015xja,Ebert:1999ht,Preis:2010cq}. Because the oscillations are not strictly coincident with each other, the size of the existing region for the SM phase also oscillates with $B$ and is the smallest around $\sqrt{eB}=0.33~\text{GeV}$ which is accidently the same as the mass in vacuum. For larger $B$, the region for the SM phase increases with $B$ due to the catalysis effect.

\begin{figure}[!htb]
\centering
\includegraphics[width=0.45\textwidth]{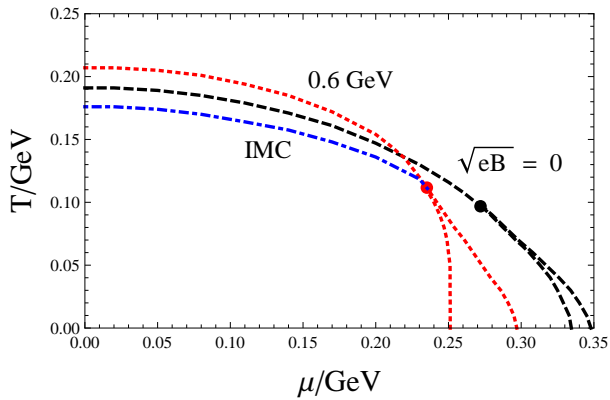}
\caption{(color online) The phase transition lines for the cases with $B=0$ (black dashed lines) and $\sqrt{eB}=0.6~\text{GeV}$ (red dotted lines). The solitonic modulation phase is in the small closed regions at large chemical potential and the bullets are the Lifshitz points. The inverse magnetic catalysis effect to the case with $\sqrt{eB}=0.6~\text{GeV}$ is also illuminated (blue dot-dashed line).}\label{Tmu}
\end{figure}

Then, we'd like to turn on the temperature and check how would the magnetic field affect the $T$--$\mu$ phase diagram. To show the effect obviously, we choose a strong magnetic field $\sqrt{eB}=0.6~\text{GeV}$ which is near the minima of the boundaries and compare it with the case in the absence of magnetic field. The results are illuminated in Fig.\ref{Tmu}.
As can be seen, the result for $B=0$ is qualitatively consistent with that obtained in Ref.~\cite{Nickel:2009wj}. However, one should notice here that a most recent study of a more general form of inhomogeneous order parameter by using finite-mode approach gives a very different result: the homogeneous-inhomogeneous phase transition becomes of first order and the inhomogeneous region is enlarged a lot even for a moderate constitute quark mass~\cite{Heinz:2015lua}. Due to the dHvA effect at zero temperature, the second-order transition lines from $\chi SB$ phase to $\chi SR$ phase intersect with each other. It might not be the case in real QCD system: Because of the IMC effect~\cite{Bali2012}, it is much more probable that the transition line would be substituted by another one for $\sqrt{eB}=0.6~\text{GeV}$ (see the blue dot-dashed line) and the transition lines are covered by those of $B=0$. However, for larger magnetic field ($\sqrt{eB}\ge 0.8~\text{GeV}$), the transition lines would start to intersect with those of $B=0$ due to the dHvA oscillation at low temperature. There are two tricritical Lifshitz points where three second-order transition lines intersect with each other in the plot and the point is found to shift to upper left by the magnetic field.

\begin{figure}[!htb]
\includegraphics[width=0.48\textwidth]{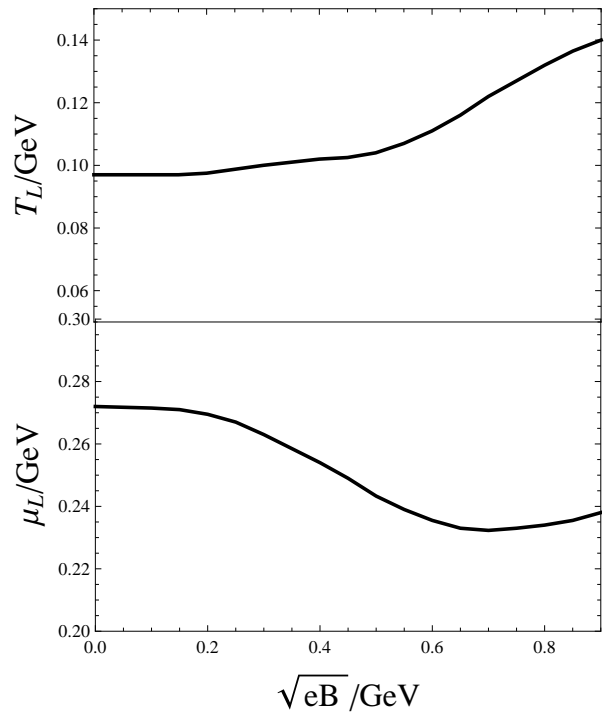}
\caption{The evolutions of the temperature $T_{L}$ and the chemical potential $\mu_{L}$ at the Lifshitz point with the magnetic field $B$.}\label{Lifshitz}
\end{figure}

Finally, the evolution of the Lifshitz point with the magnetic field is evaluated as illuminated in Fig.\ref{Lifshitz}.
The dHvA oscillation shows up in both curves: The oscillation is much more obvious in the $\mu_L$--$B$ curve with the minimum at around $\sqrt{eB}=0.7~\text{GeV}$. On the other hand, the temperature $T_{L}$ increases monotonically with the magnetic field $B$ and the flat oscillation in the region $\sqrt{eB}\in[0.2~\text{GeV},0.5~\text{GeV}]$ is mainly due to the fast decreasing of $\mu_L$.

\section{Summary}\label{summary}
In this work, we explore the magnetic field effect to the phase transitions among the homogeneous chiral symmetry breaking,  inhomogeneous solitonic modulation and chiral symmetry restoration phases. The thermodynamic potential can be obtained directly by generalizing from the case without magnetic field due to the sign symmetry of the quark spectra. And in order to evaluate the chiral symmetry restoration transition more efficiently, we develop a new kind of Ginzburg-Landau expansion approach around small $\nu$.

The transitions from $\chi SB$ phase to SM phase and from SM phase to $\chi SR$ phase are both of second order in the presence of magnetic field. At zero temperature, both the upper and lower boundaries of the SM phase oscillate with the magnetic field and so does the size of the SM region which are illuminations of the dHvA effect. At finite temperature, the tricritical Lifshitz point is also found to oscillate with the magnetic field. Generally, only one minimum is found for each $\mu$--$B$ phase diagram which is consistent with previous works~\cite{Cao:2015xja,Ebert:1999ht,Preis:2010cq}.

Our recent study is very preliminary. How's the fate of the SM phase when the fluctuations of collective modes are included either through Gaussian expansion (though very difficult) or quark-meson model~\cite{Fraga:2008qn,Nickel:2009wj,Carignano:2014jla} is an important question. As it is impossible to check which one is much more favored between the DCDW modulation and the solitonic modulation phases by the first principal LQCD calculation at finite chemical potential~\cite{Nakamura:1984uz}, it is very important to find a method that can consistently treat both cases at finite magnetic field. The Dyson-Schwinger equation and the functional renormalization group approach may serve as possible candidates. Still, when taking a more general form of inhomogeneous order parameter into account~\cite{Heinz:2015lua}, how would the phase diagram change at finite magnetic field deserves further study.  Finally, though the dHvA oscillation surely exists at low temperature, how would the IMC effect affect the $T$--$\mu$ phase diagram at finite magnetic field still need further check.

{\bf Acknowledgments:} G.C. thanks Bernd-Jochen Schaefer for providing the relevant papers about the study of inhomogeneous states and Lifshitz point. The work is supported by the NSFC under Grant No. 11335005 and the MOST under Grants No. 2013CB922000 and No. 2014CB845400.

\end{document}